# IMPLEMENTATION OF TINY MACHINE LEARNING MODELS ON ARDUINO 33 – BLE FOR GESTURE AND SPEECH RECOGNITION

Viswanatha V[1*]
*Asst.Professor, Electronics and Communication Engineering Department,*
*Nitte Meenakshi Institute of Technology, Bangalore.*
*Email – viswas779@gmail.com*

Ramachandra A.C[2]
*Professor and Head, Electronics and Communication Engineering Department*
*Nitte Meenakshi Institute of Technology, Bangalore. Email- ramachandra.ac@nmit.ac.in*

Raghavendra Prasanna R [3]
*BE Student, Electronics and Communication Engineering Department*
*Nitte Meenakshi Institute of Technology, Bangalore. Email- raghavendraprasanna2000@gmail.com*

Prem Chowdary Kakarla [4]
*BE Student, Electronics and Communication Engineering Department*
*Nitte Meenakshi Institute of Technology, Bangalore. Email- chowdaryprem5@gmail.com*

Viveka Simha PJ [5]
*BE Student, Electronics and Communication Engineering Department*
*Nitte Meenakshi Institute of Technology, Bangalore. Email- vivekasimhapj@gmail.com*

Nishant Mohan [6]
*BE Student, Electronics and Communication Engineering Department*
*Nitte Meenakshi Institute of Technology, Bangalore. Email- nishu1926@gmail.com*

[*] *Corresponding Author:* Viswanatha V

**Abstract-**

**In this article gesture recognition and speech recognition applications are implemented on embedded systems with Tiny Machine Learning (TinyML).The main benefit of using TinyML is its portability. It can be run on cheap microcontrollers with tiny batteries and low power consumption, and it can easily integrate machine learning with virtually anything. It also has the benefit of increased security due to local nature of computing. The benefit of using Arduino Nano 33 BLE sense is that it has a set of sensors embedded on the top, which gives us a lot of options to try ideas without having to generate the circuit to such sensors in prototyping board. It features 3-axis accelerometer, 3-axis gyroscope and 3-axis magnetometer.**

**The gesture recognition, provides an innovative approach nonverbal communication. It has wide applications in human-computer interaction and sign language. Here in the implementation of hand gesture recognition, TinyML model is trained and deployed from EdgeImpulse framework for hand gesture recognition and based on the hand movements, Arduino Nano 33 BLE device having 6- axis IMU can find out the direction of movement of hand.**

**The Speech is a mode of communication. Speech recognition is a way by which the statements or commands of human speech is understood by the computer which reacts accordingly. The main aim of speech recognition is to achieve communication between man and machine. Here in the implementation of speech recognition, TinyML model is trained and deployed from EdgeImpulse framework for speech recognition and based on the keyword pronounced by human, Arduino Nano 33 BLE device having built-in microphone can make an RGB LED glow like red, green or blue based on keyword pronounced. The results of each application are obtained and listed in the results section and given the analysis upon the results.**

**Keywords: TinyML , Gesture recognition ,Speech recognition, Machine Learning , Hardware implementation**

## I. INTRODUCTION

We experience a daily reality such that AI models assume a significant part in our day to day existence. Everyday undertakings like snapping a photo, checking the climate and so forth rely upon AI models yet preparing the model and running connection points are costly. Little ML calculations generally work the same way as a normal AI calculation. The models are prepared on the cloud or a client's PC. Subsequent to preparing, undertakings become possibly the most important factor in cycles of model pressure. It's a field of learning in machine learning systems and embedded programs that explores such models you can use on nearly nothing, less solid contraptions as little controls. Enables low inactivity, low power and low exchange speed model for contraptions. Regularly, Tiny ML grants IOT based embedded edge contraptions to go to cut down power structures with combination of refined power the board modules[1]-[5]. In the gathering region, Tiny ML can stop edge time in view of stuff dissatisfaction by engaging consistent decision. ML running at embedded edge devices as shown in fig.1 results in low processing latency, better privacy and minimal connectivity dependency[6]-[10].

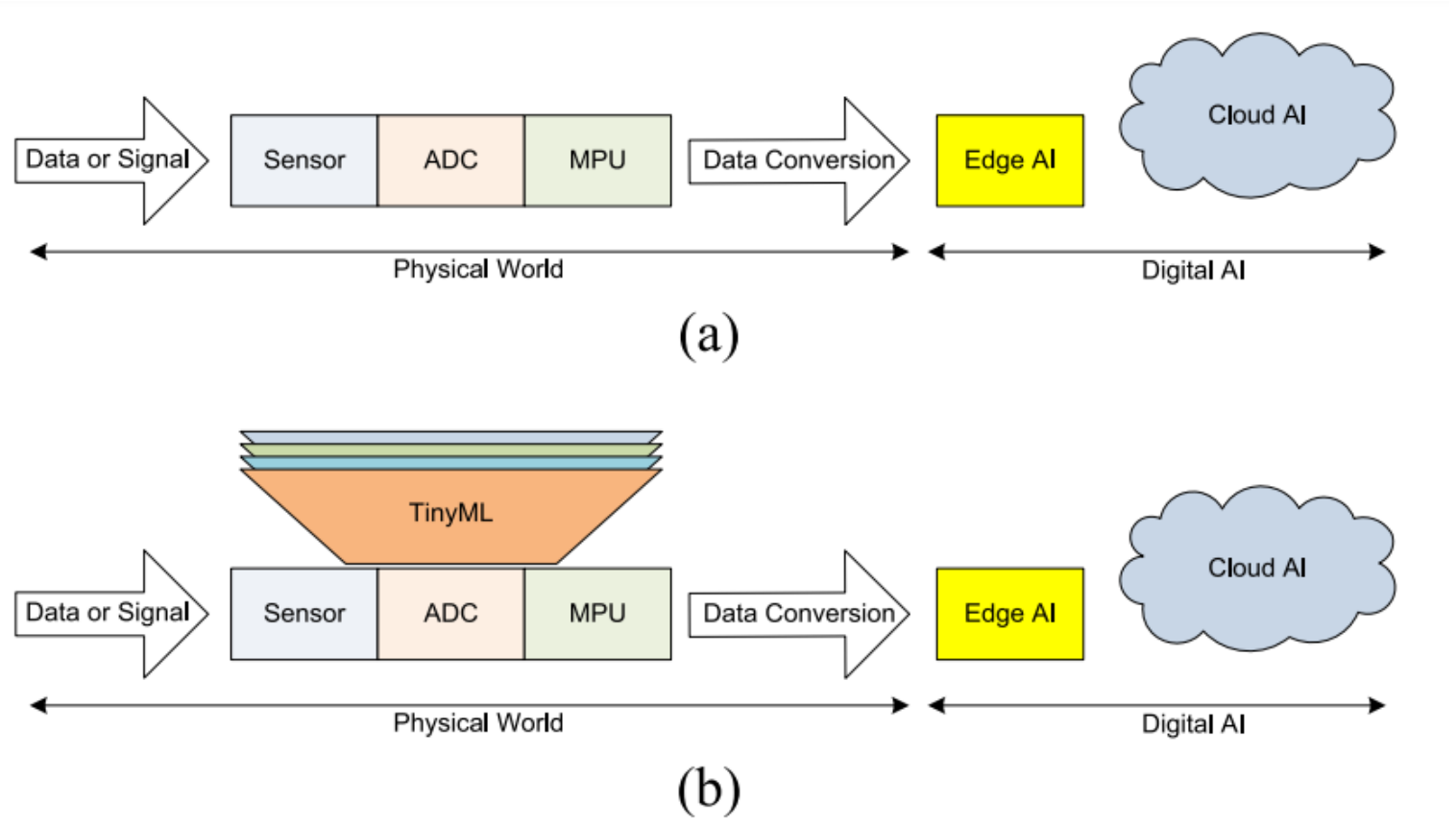

Figure.1 (a) Physical world and digital AI, (b) Tiny ML assisted digital AI

Fig.2 represents technology required at cloud ML, edge ML and TinyML .The technology is in terms of algorithms, hardware and so on. TinyML technology takes data from sensors and give it to the micro or nano level convolution neural network where microcontrollers are used to run the neural networks. Such microcontrollers may have hardware accelerators. In case process is so complex then such process can be taken into deep neural network with the help of GPU, multi-core CPUs and TPU. Fig.3 represents layered approach implementing ML [11]-[16].

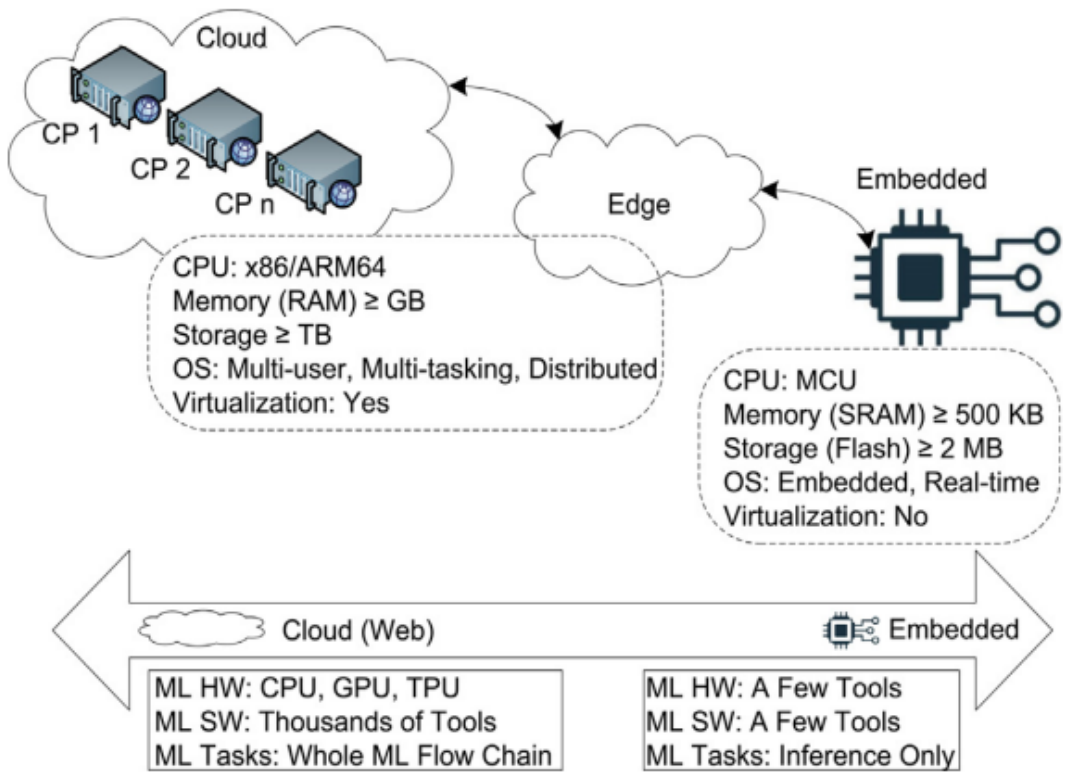

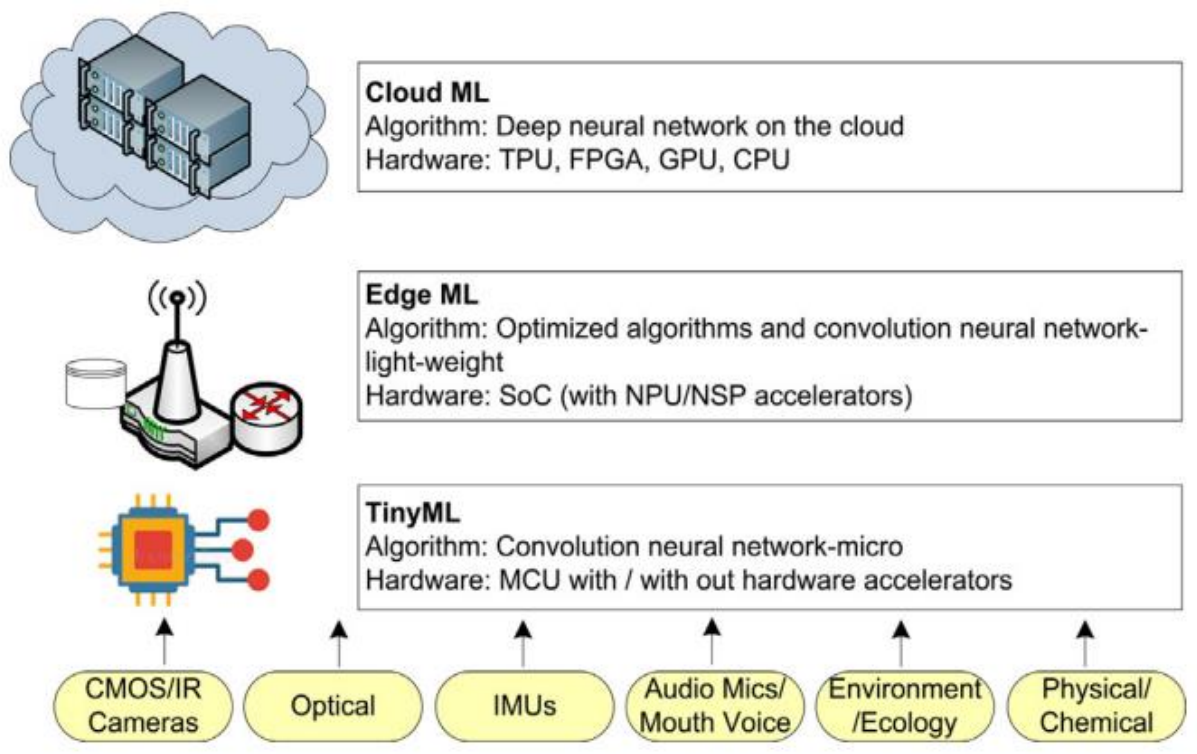

Figure. 2 Miniscule Scalability of Tiny ML.

Figure.3 TinyML with layered approach

To support configuration, screen and adjust the small ML application, we should utilize "ML capabilities". ML occupations accomplish basically everything consequently. According to the viewpoint of all stages, they oversee and handle information, train AI models, convert them, test them, look at them, and use them. Countless devices might be essential for the environment. Likewise, minuscule ML as a help that engages creation locales to effectively oversee and coordinate different little ML gadgets. A profoundly scattered environment might develop in light of the fact that little ML implanted gadgets, from ML integrators and ML processing frameworks, are intended to accomplish very low power productivity. At the point when the equipment changes, the split influences the progression of the coordinated ML model [17]-[20]. Tiny ML frameworks can take direct information input from different sensors. It can utilize a miniature and Nano level convolution brain organization.

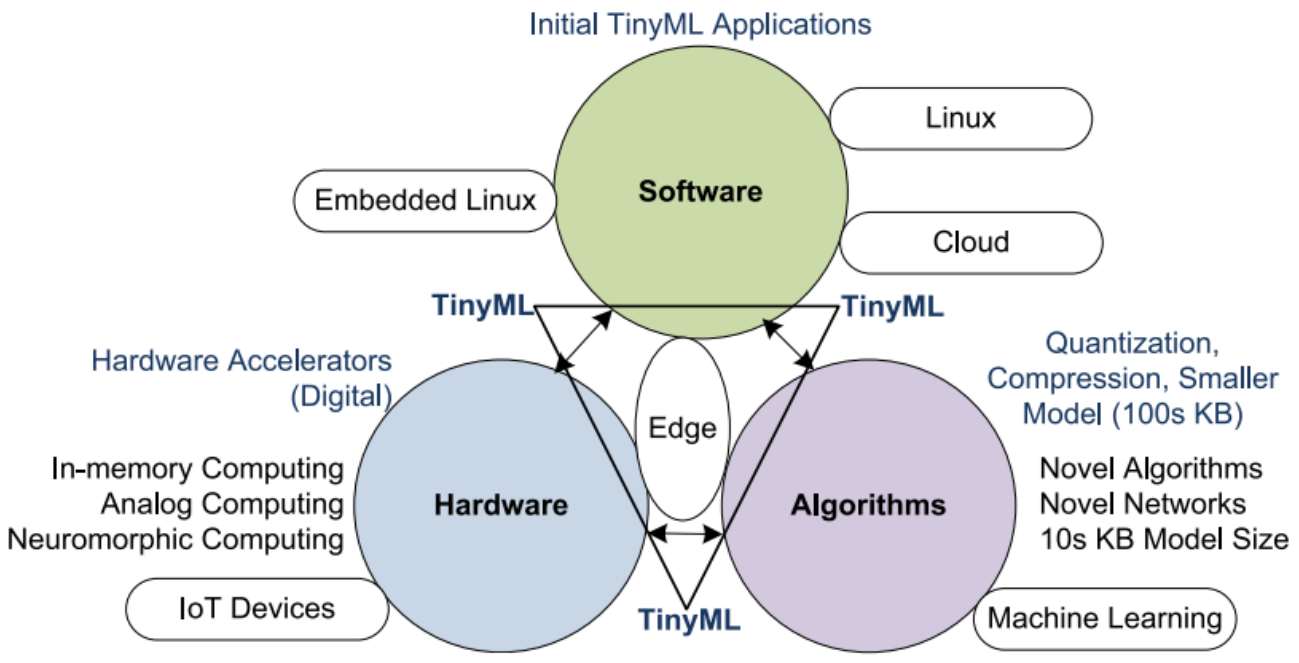

Figure. 4 Composition of Tiny ML.

The fig.4 presents key parts of the Tiny ML where the total mix of equipment programming co-plan is the main variable. Such frameworks ought to go past the AI bend gave top notch information and incorporated programming plan. Ordinarily, the Tiny ML framework is enlightened by double documents created from a prepared model on a huge facilitating machine [21]-[24].

There are certain Programming and software libraries available for TinyML Implementation ; *Tensor Flow Lite (TFL):* It is a top to bottom perusing system with open source help for an educated understanding idea. Edge-empowered AI on the gadget can be settled by this structure while utilizing five key hindrances (e.g., delays, protection, network, size, and power utilization). Upholds Android, iOS, installed Linux, and an assortment of microcontroller. It additionally upholds dialects (e.g., C ++, Python, Java, Swift, C) to further develop AI on the fringe gadget.
*Tensor:* It is a free implanted learning climate that makes model and moment sending on IOT-edge gadgets. Tensor is a little size module that requires just 2 KB of circle.
*Edge Impulse:* It is a cloud administration for creating AI models on Tiny ML-coordinated edge gadgets. This supports Auto ML handling for start to finish areas. It likewise upholds various sheets that coordinate advanced cells to create learning models on such gadgets. Preparing is finished on the cloud stage and a prepared model can be conveyed on the fringe gadget as per the empowered information move technique. The force can be run on a nearby machine with the assistance of the underlying C ++, Node.js, Python, and Go SDK.
*Nano Edge AI Studio:* The product was previously known as Cartesiam.ai, presently it considers the choice of the best library and test library usefulness by utilizing an emulator before the last edge.
PyTorch Mobile: It is important for the PyTorch environment that plans to help all stages from preparing to the utilization of AI models to advanced cells (e.g., Android, iOS). A couple of APIs are accessible to pre-process AI in versatile applications
*Installed Reading Library (ELL):* Microsoft has created ELL to help the Tiny ML environment for inserted perusing. Offers help for Raspberry Pi, Arduino, and miniature: piece stages. Models utilized on such gadgets are undetectable web, so no cloud access is required. Upholds picture and sound division at present.
*STM32Cube.AI:* It is a code generator and improvement programming that permits AI and AI-related errands to be rearranged on STM32 ARM Cortex Mbased (STM32Cube.AI, 2021) sheets. The utilization of brain networks on the STM32 board can be accomplished straight by utilizing the STM32Cube.AI to change over brain networks into the most reasonable MCU code.

## II. DESIGN APPROACH AND METHODOLOGY

Hardware board used in this development is Arduino Nano 33 BLE as shown in fig.5 is progressed PC stages for utilizing tense AI models. It contains a 32-bit ARM Cortex-M4F microcontroller running at 64MHz with 1MB of framework memory and 256KB Smash. This little regulator gives sufficient ability to utilize Tiny ML models.

The Arduino Nano 33 BLE Sense contains variety, splendor, closeness, contact, development, vibration and different sensors. This sensor suite will be all that could possibly be needed for most applications

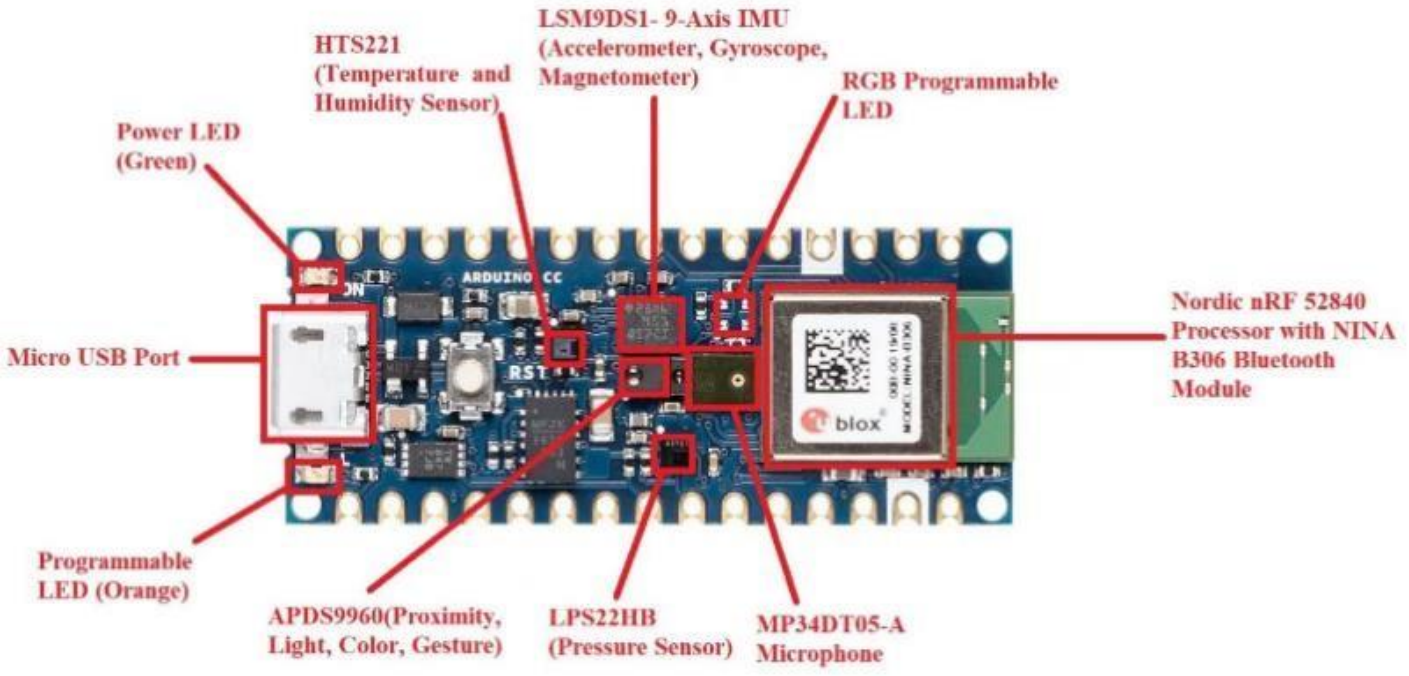

Figure .5 Arduino Nano 33- BLE

Machine learning framework as shown in fig.6 represents model of EdgeImpulse.
There are a couple of structures that address the issues of TinyML. In that sense, the EdgeImpulse is extremely famous and has extraordinary social help. Utilizing edge drive, it can convey models on microcontroller.

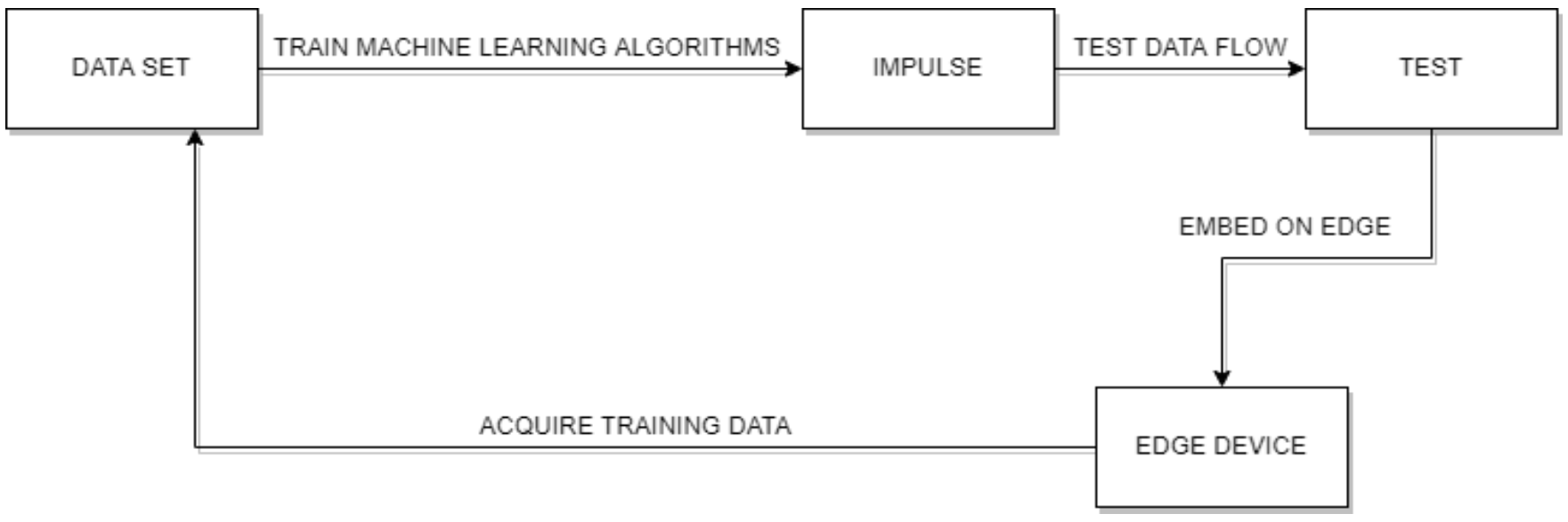

Figure. 6 Block diagram of Edge Impulse

The methodology followed in the implementation of both the applications is as shown in fig.7
.

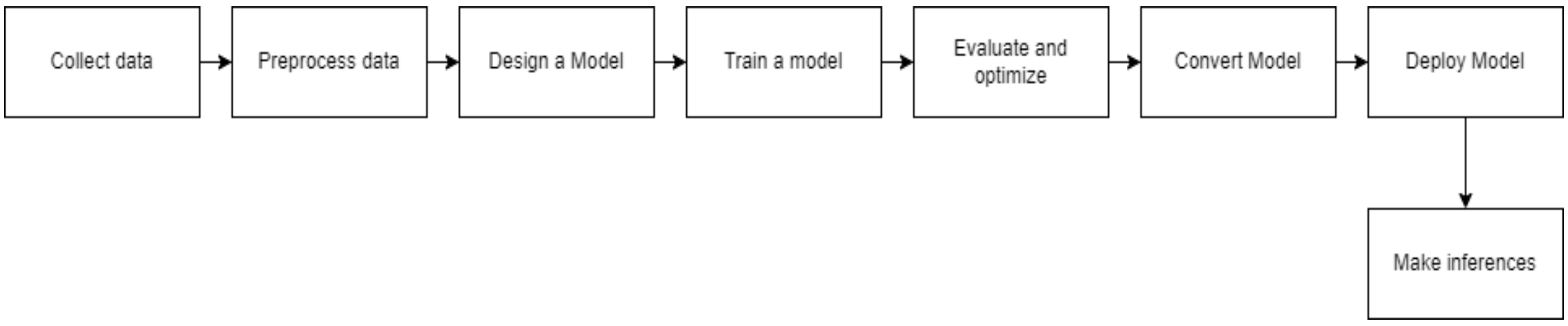

Figure .7 Block diagram of flow of implementation of TinyML

Steps to work on edge impulse:
To start, first make a record by going to https://studio.edgeimpulse.com/information exchange. Subsequent to entering in your data and checking your email, you will be welcomed by a getting everything rolling page. This will walk you through the most common way of interfacing a gadget, gathering information, lastly conveying a model. I named my most memorable task PhoneTest-1, yet it tends to be anything you like.
Interfacing a Cell Phone: Tiny ML upholds numerous gadgets, including the ESP32, numerous ST ARM Cortex-M3 sheets, and a few Arduino Wi-Fi-empowered units. Nonetheless, a large number of similar errands can be achieved just by utilizing a cell phone by means of an internet browser, as it contains a receiver and accelerometer. To interface your telephone, just snap on the "Utilization your cell phone" button which opens up a QR code. In the wake of checking it, you will be taken to their site and consequently associated with their Programming interface through a Programming interface key. Make a point to keep your telephone on and the program window open all through the remainder of the aide.
Gathering Information: Presently it is the right time to make a plunge and make a model, as a matter of fact. On the whole, there must be information to prepare it on. Ensure you have your telephone convenient on the grounds that you'll utilize its sensors to catch the information. To start, go to the information procurement tab and ensure your telephone is chosen. Pick the accelerometer sensor and recurrence and afterward click "Begin examining." After you are finished moving your telephone, you can see the gathered information in a diagram.
Preparing a Model: Since you have a few information recorded, now is the ideal time to prepare a model from it. Feel free to explore to the "Make Motivation" page and select the suggested Ghastly Investigation handling block and a Keras Brain Organization learning block. Then feel free to save the drive. Then, set up your information's scaling, channel, and FFT settings. These will control how your information gets pre-handled prior to being sent into the NN. In the wake of doing that, view and create the highlights. On the NN settings page, I chose to change the default certainty edge from 80% up to 91%. In the wake of preparing the model, I had the option to see a diagram of what the model thought of. Then, at that point, I went to the "Order" page and accumulated a smidgen more information from my telephone and saw what the model had the option to recognize.
Sending: To send the model, I sent out the model as a Web ASM record and afterward unfastened it. Then another js record called run-impulse.js and put it into a similar organizer as the model (however the document is connected to this task page). To run it, I entered the hub order into the order brief followed by run-impulse.js and afterward glued the "Crude elements" cluster in statements as the second contention for the hub order.

There are two applications implemented in this article. i). Gesture recognition to find the movement of human hand. ii) Speech recognition to control on board RGB LED of Arduino 33 BLE board.This incorporates both an Arduino IDE and EdgeImpulse framework .EdgeImpulse frame works has built-in tools, libraries and nxn algorithms to build the model .In EdgeImpulse tool, the process starts with sampling to generate the data called as training dataset. Later test dataset is generated by split process. Training data and test data will be in the ratio of 80/20. Pre-processing is carried out on the data using spectral analysis. Later classification using NN classifier is carried out using Keras method .The next step is to feature extraction which is done on training dataset followed by classification. Last step is to test the model (if the results are not satisfied, retrain the model) followed by build the model for deployment into the hardware. Fig. 8 shows the flowchart for gesture recognition to find the movement of human hand and fig.9 shows the flowchart for speech recognition to control RGB LED.

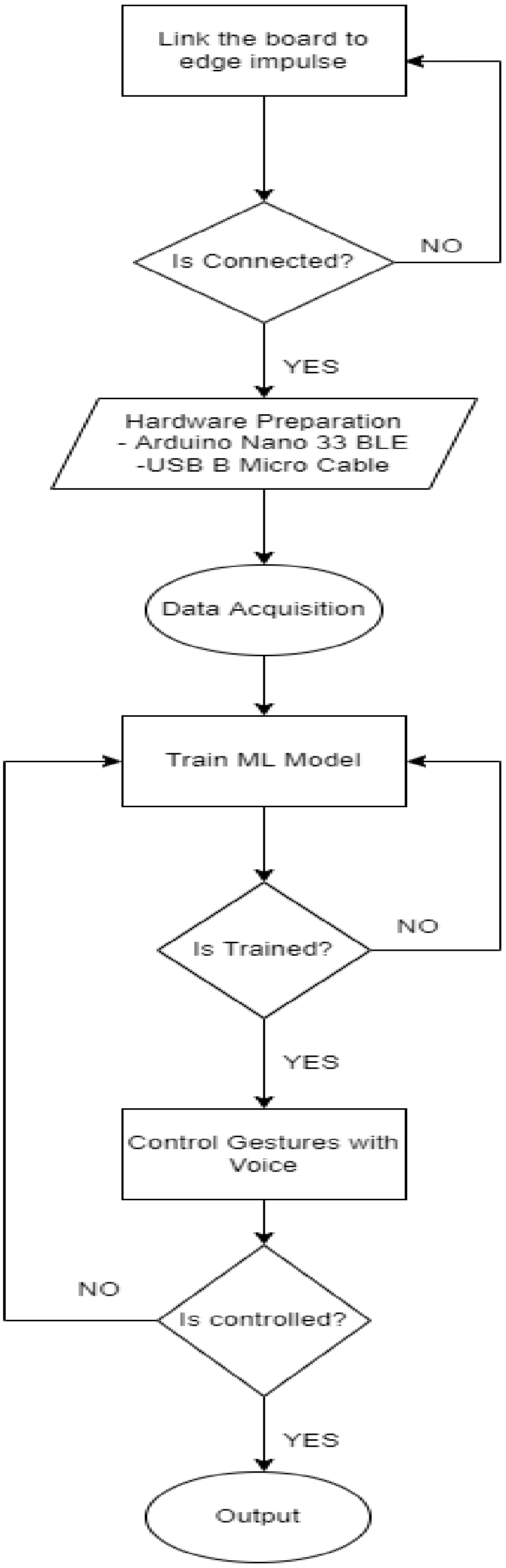

Figure.8 Flowchart for hand gesture recognition

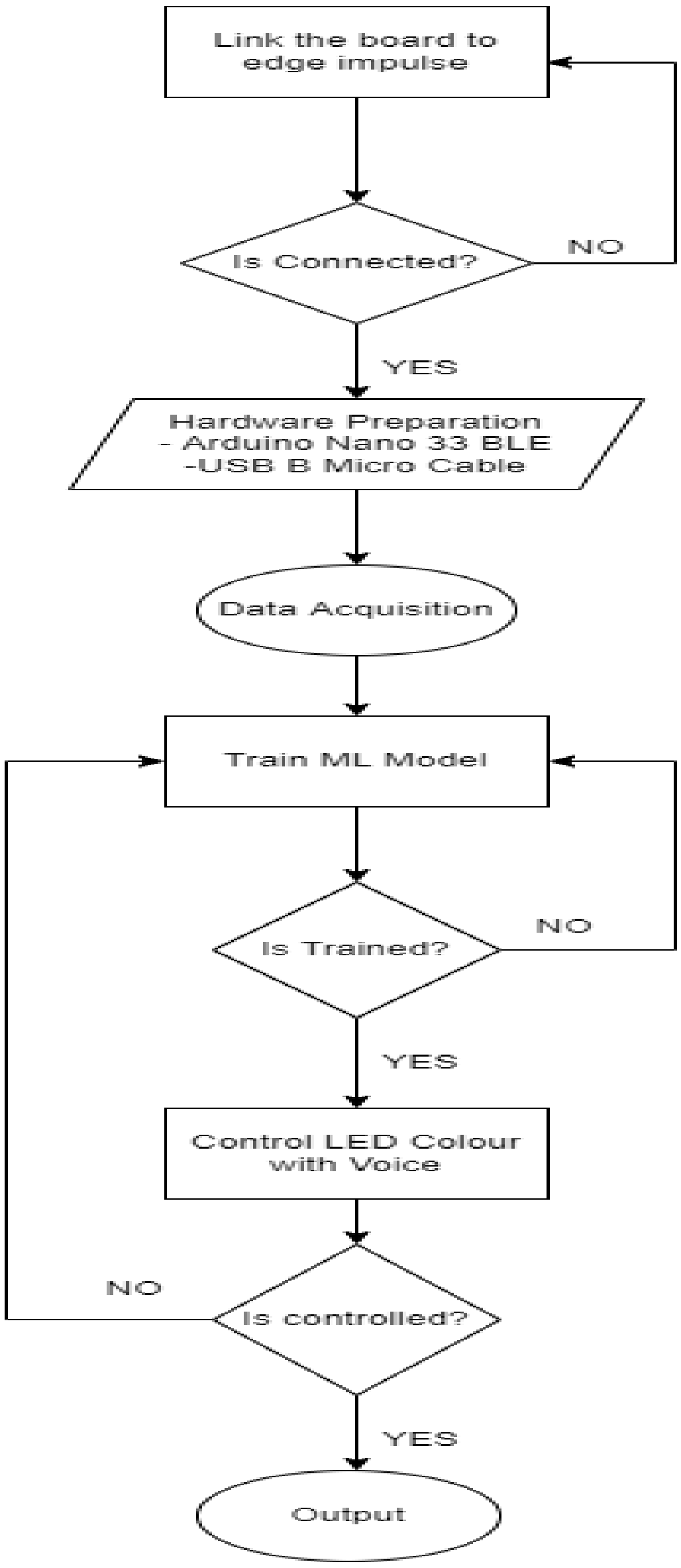

Figure.9 Flowchart for speech recognition

## III. RESULTS AND DISCUSSION

### A). Hand Gesture Recognition

- Get started with Machine Learning as shown in fig.10
- Login to Edge Impulse
- Create an undertaking
- Select the Arduino board in project data

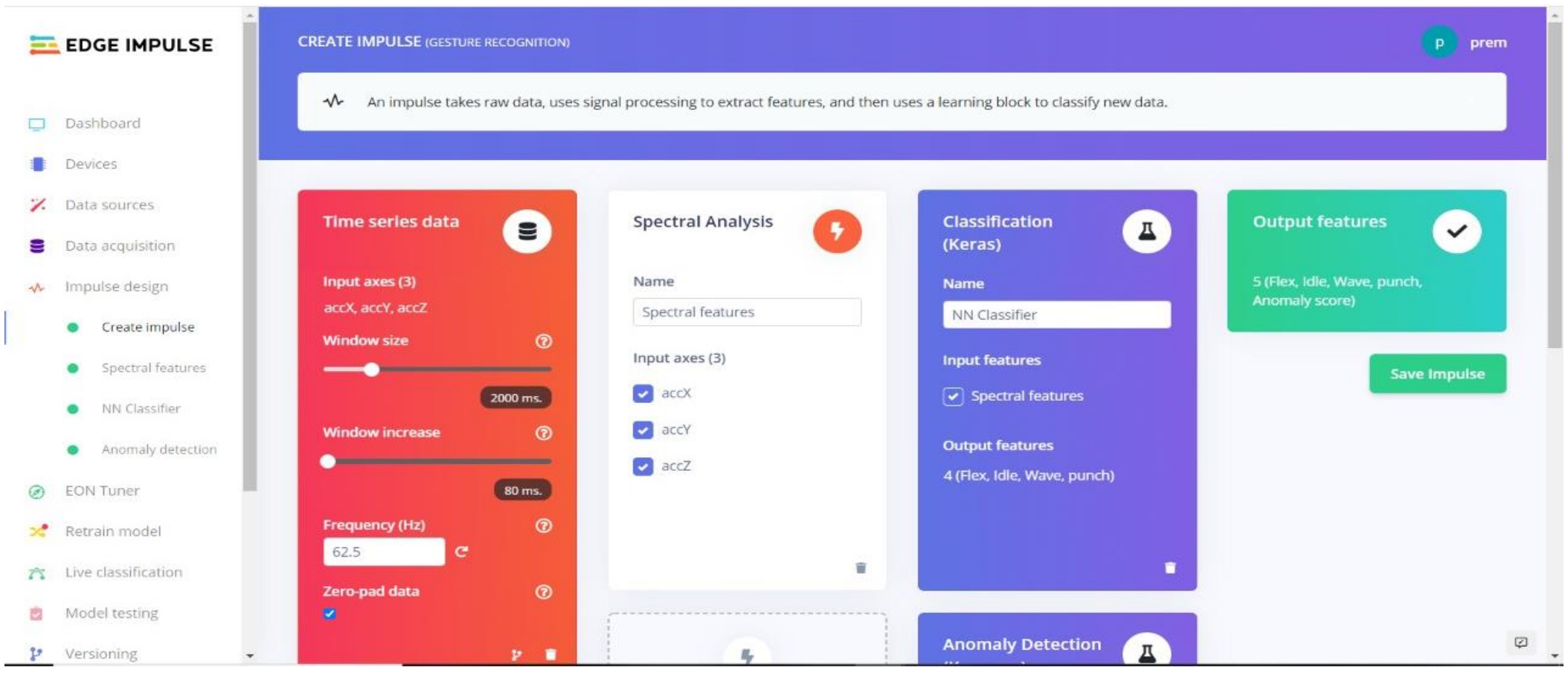

Figure. 10 Impulse design

Train ML model

- After the information is gathered go to "Impluse Plan.
- Click on Add handling block and select spectral analyser.
- Click on Add learning block and select order and Save Impluse.
- Click on SPA, use Default setting and snap on Save Boundaries.
- Click on Produce Elements and trust that handling will wrap up.
- To train information click on NN Classifier and classification results are available as shown in fig.11 with training dataset.

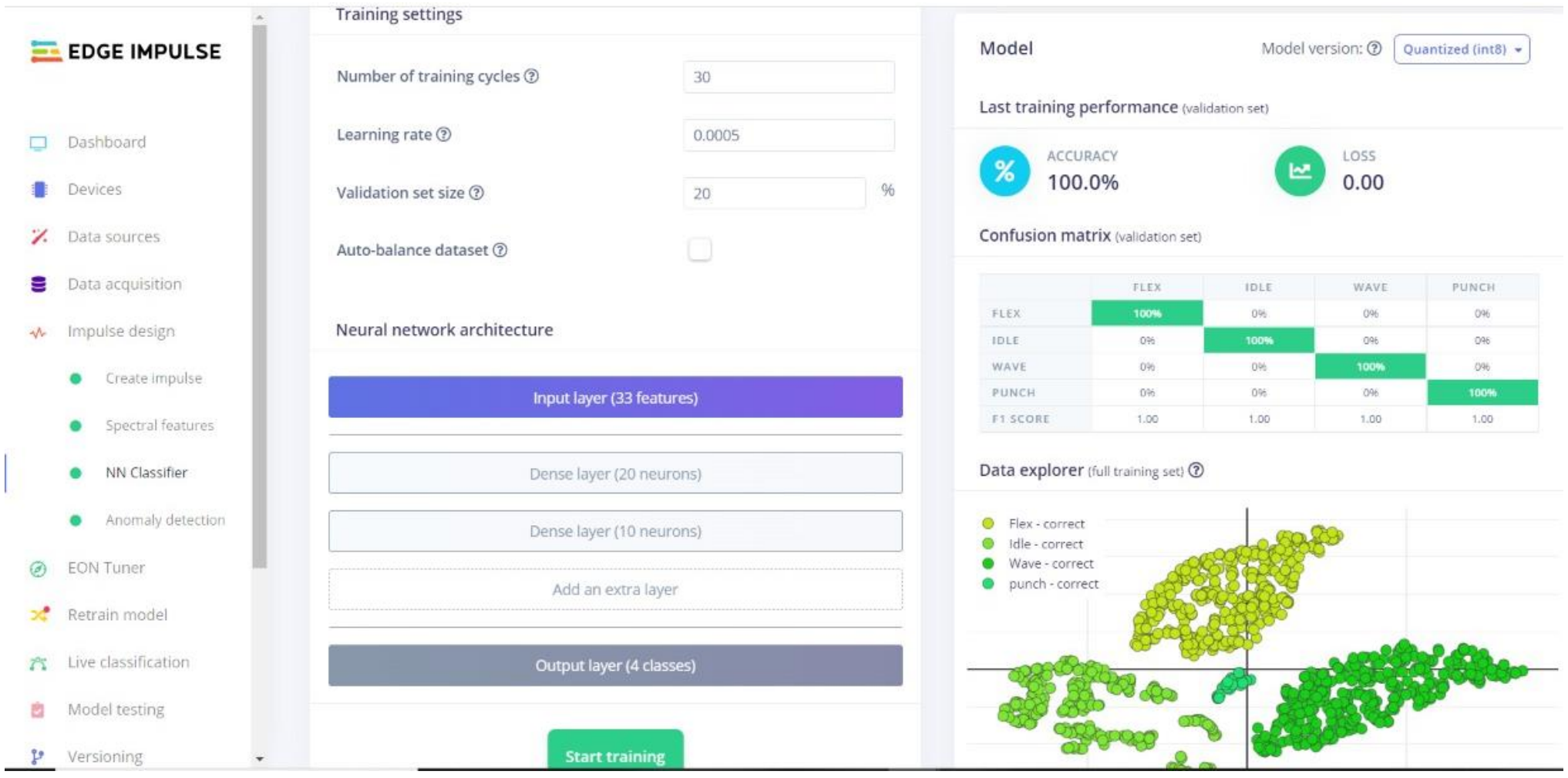

Figure. 11 Training data

- Tick on "Information Expansion" and snap on Begin Preparing.
- Ensure that there is a precision above 90%.
- Go to "Model Testing”, Set least certainty rating to 0.6 and click on "Arrange all" to obtain test dataset.

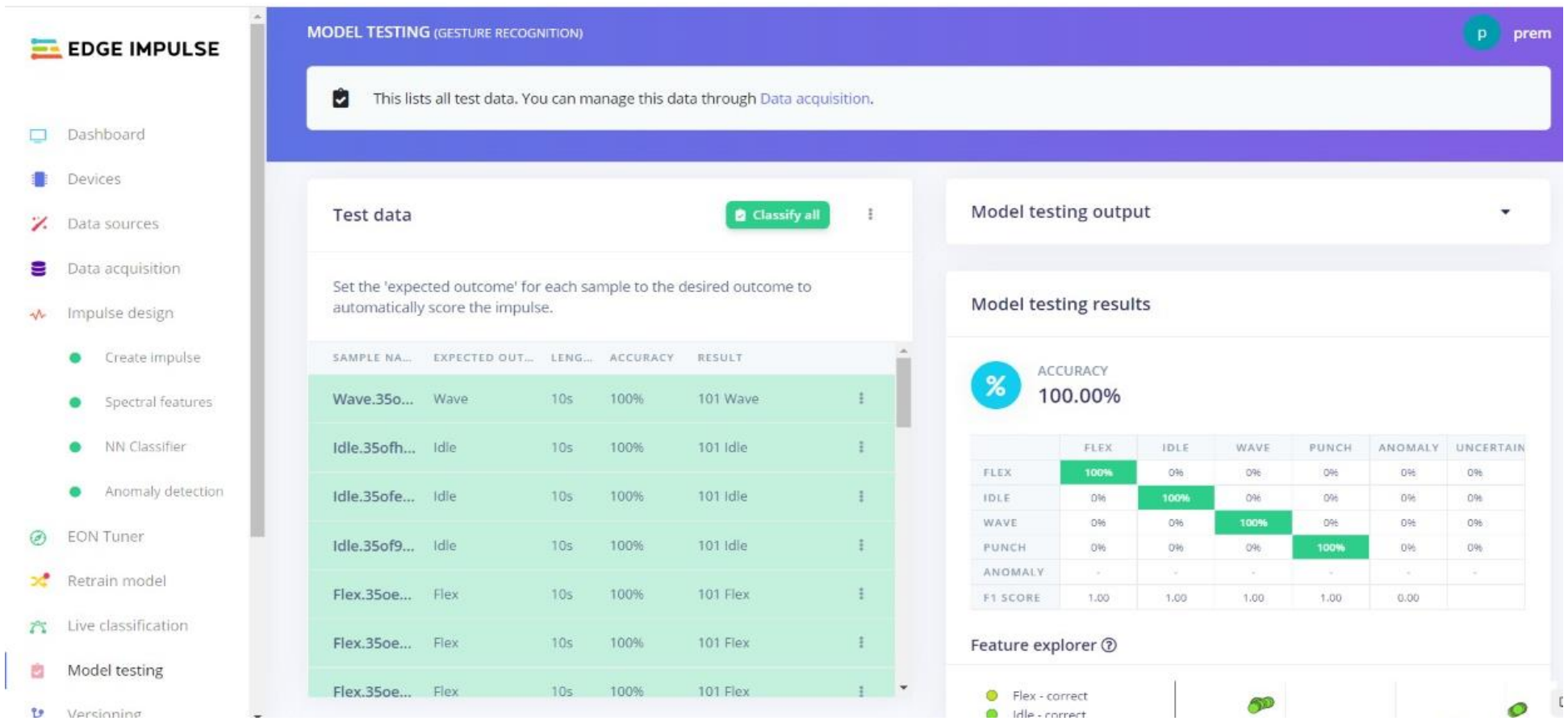

Figure .12 Test data

- Ensure there is an outcome with great exactness.
- Select Organization, click on Arduino library and snap on Form.
- We get a Compress record which can be utilized on Arduino IDE for deployment as shown in fig.13.
- Movement of hand with Wave and flex movement is detected by the model after deployed into Arduino 33 BLE as shown in fig.14.

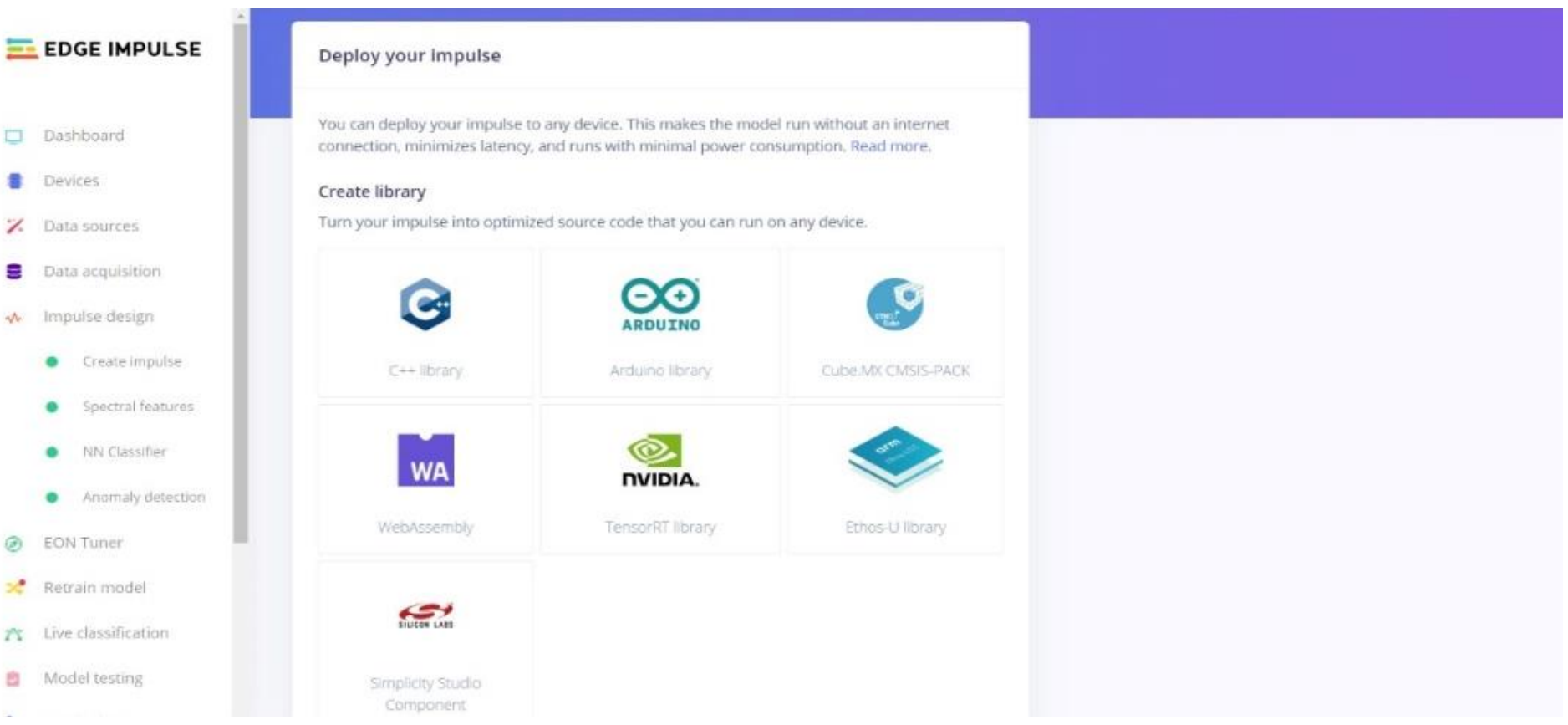

Figure .13 Deployment

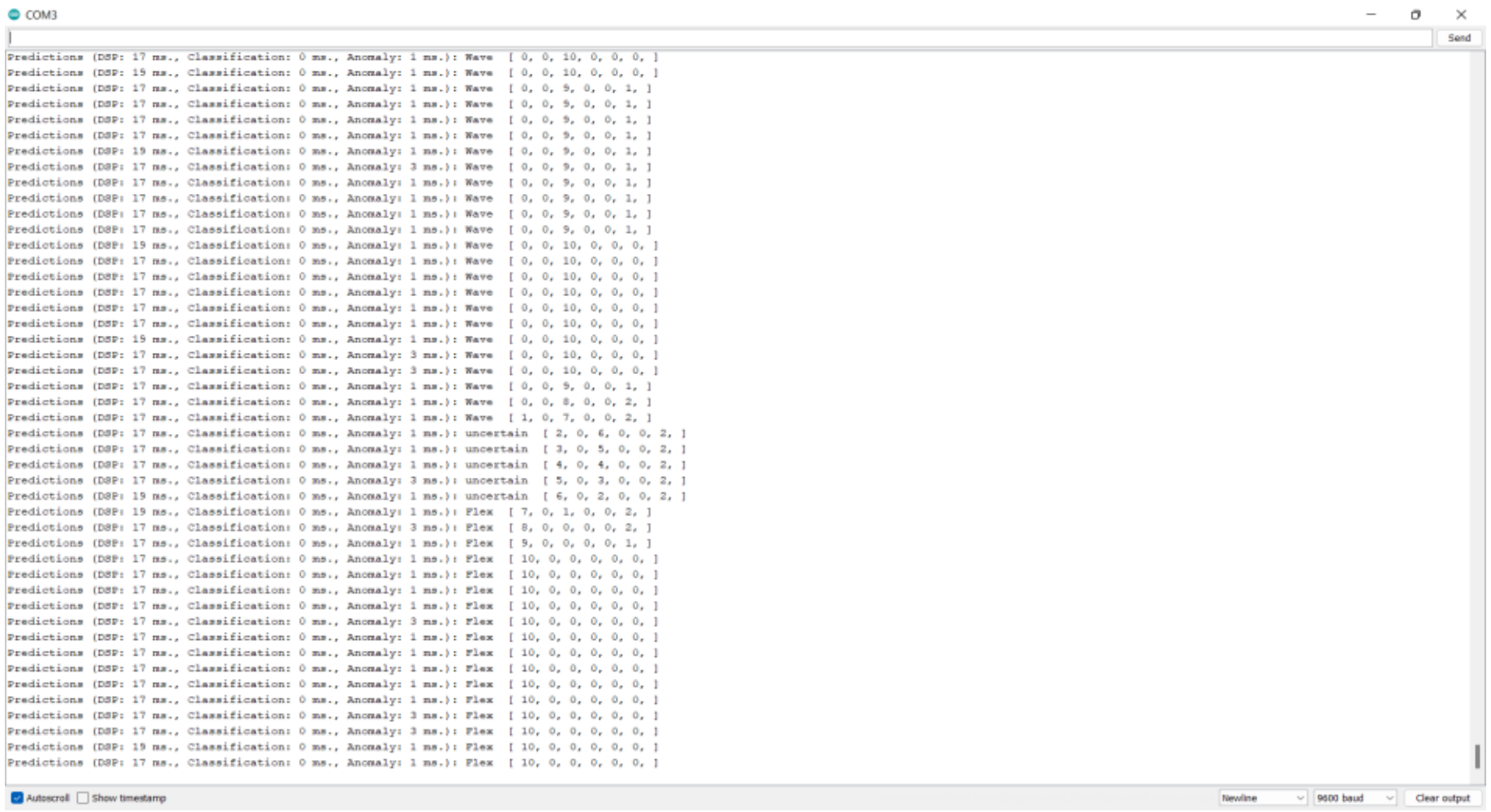

Figure .14 Resultant Output. Wave and Flex movement identification

B). Speech Recognition. Same procedure is followed for this application too Therefore it gives the results as shown fig.15 for model testing.

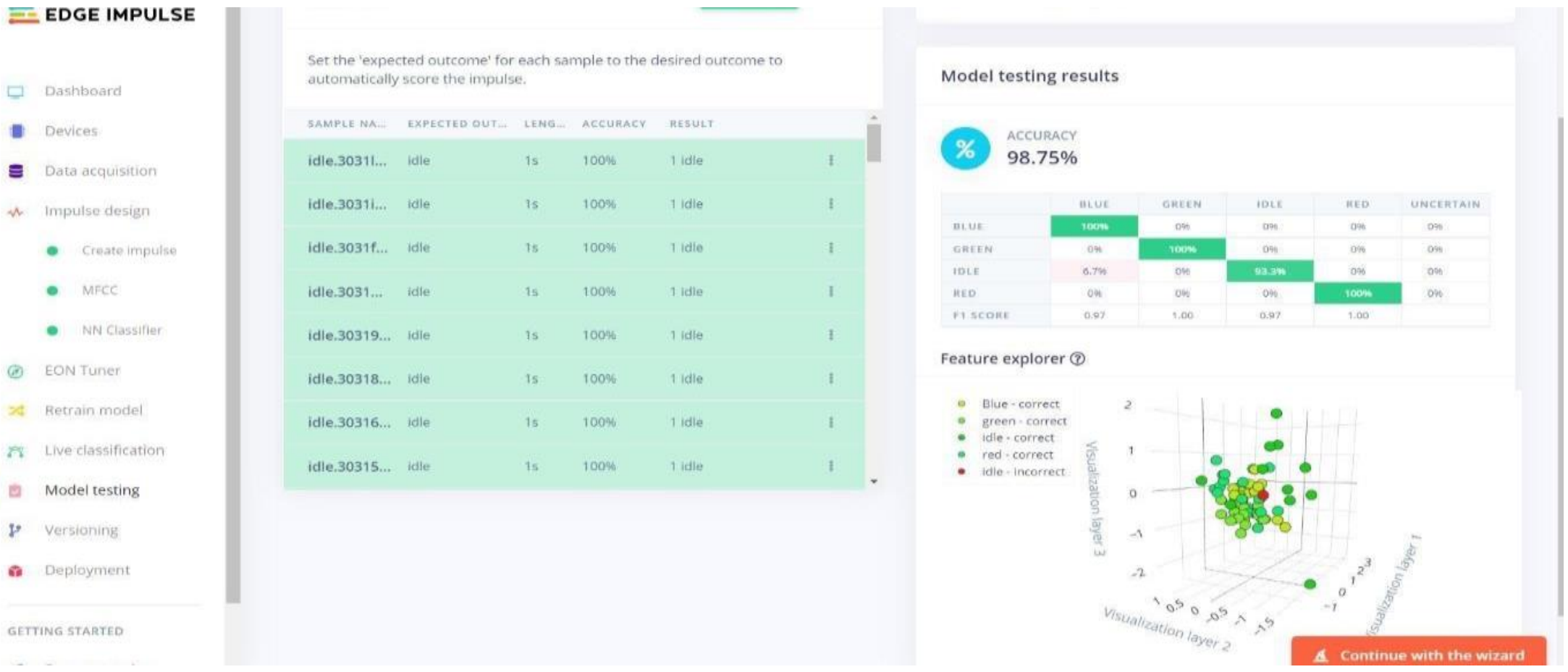

Figure. 15 Model testing

Control LED with Voice

- Add the Compress record we acquired from EdgeImpluse into Arduino Library.
- Open the mouthpiece ceaseless model and Select the right board and port.
- Run and Transfer the program ready and Open Chronic screen.
- It will show expectation for tones and inactive state, when a variety is called its forecast worth will Increment
- We can alter the code to control worked in RGB Drove ready.
- In fabricated Drove will shine Red when Red is called as shown in fig.16

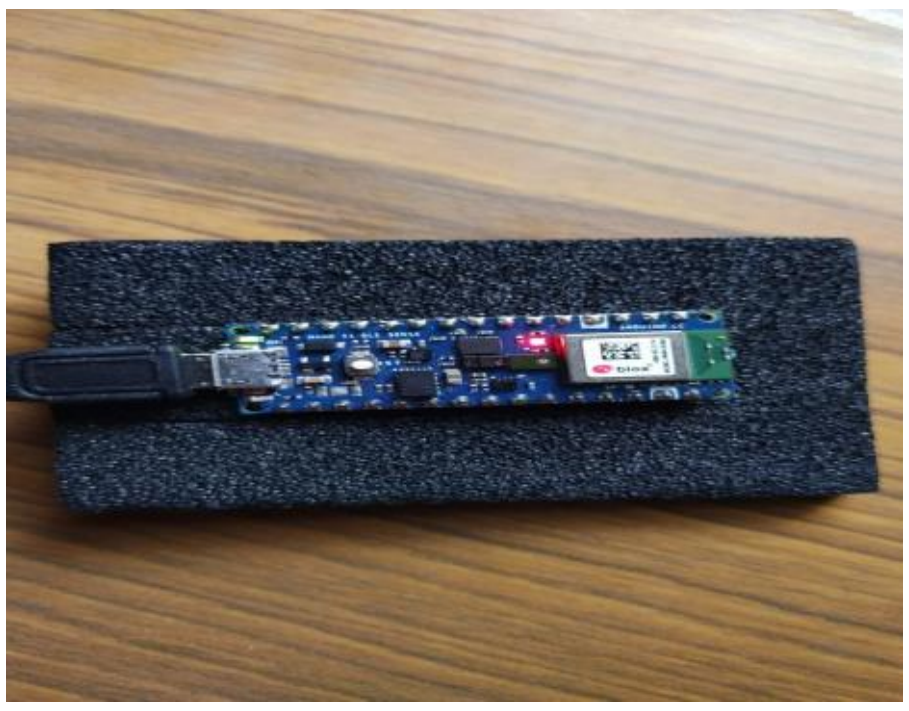

Figure. 16 Output

## IV. CONCLUSION AND FUTURE SCOPE

The hand signal acknowledgment framework that is intended to have the option to perceive hand motions progressively, this is an image of the looking at a climate where a fashioner can portray by controlling pointer utilizing a couple of digital gloves and can cooperate with the plan item in 3D space. Speech recognition framework will be all the more broadly utilized from now on. An assortment of discourse acknowledgment items would show up on the lookout. It is hard to deliver a discourse acknowledgment framework work precisely like a human. Presently the discourse acknowledgment innovation must be acquainted into individuals' lives with bring more accommodation. Specialists accept discourse acknowledgment is one of the vital impending advancements in the field of data.

Later on, different hand motions can be perceived and applied as contribution to the PC. The hand motions addressing numbers can likewise be changed over into orders to perform related undertakings continuously.